\documentclass[11pt]{article}
\usepackage{moriond,epsfig}

\def\be{\begin{equation}}
\def\ee{\end{equation}}
\def\bea{\begin{eqnarray}}
\def\eea{\end{eqnarray}}

\begin{document}
\vspace*{4cm}
\begin{center}
  {{\large{\bf REGGE DESCRIPTION OF $\pi\pi$, $\pi K$, $\pi N$,$K N$ AND $NN$ 
FORWARD SCATTERING ABOVE $E_{kin}>1\,\hbox{GeV}$}}}
\end{center}


\vskip .5cm
\centerline{ \textsc{J.R. Pel\'aez$^1*$}}

\begin{center}
$^1$Departamento de F\'{\i}sica Te\'orica,~II,
Facultad de Ciencias F\'{\i}sicas,\\
Universidad Complutense de Madrid,
E-28040, Madrid, Spain.
\end{center}


\vskip 1cm

\begin{center}
  \begin{minipage}[h]{13cm} \small
We review our recent description of high energy forward hadronic cross sections
and in particular of
pion scattering data by means of Regge theory. 
We obtain a prediction for the total $pp$
cross section at LHC. In addition, and contrary to
some suggestions in the literature we show that standard Regge behavior
can accommodate both crossing symmetry and factorization.
Some consequences for low energy $\pi\pi$
scattering are briefly discussed.
\end{minipage}
\end{center}


\section{Introduction}

A correct description of $\pi\pi$ scattering at high energies
is crucial to achieve a very precise dispersive evaluation of 
$\pi\pi$ low energy observables. This is of particular interest to test
Chiral Perturbation Theory (ChPT), and in particular the values of the
chiral parameters and of the chiral condensate.
Since $\pi\pi$ scattering amplitudes are very constrained by
analyticity, dispersive approaches, using data in a larger
energy range, can improve the information
from low energy data alone, which are not very reliable 
and are affected by large
systematics. 

At high energies, the formalism able to deal with hadronic cross sections
is Regge theory, which is as much part of QCD as ChPT. 
It describes amplitudes in terms of {\it Regge poles},
which are complicated objects related to the 
$t$ channel exchange of resonances, like the reggeized $\rho$ 
in the isospin 1 channel, 
or the {\it Pomeron} when no quantum numbers are exchanged.
All Regge pole contributions 
decrease at large $s$, except that of the Pomeron,
so that all $\pi\pi$ total cross sections,
tend to a common value  at sufficiently high energies ($\simeq 20\,$GeV), 
denoted $\sigma^\infty$, (Pomeranchuk Theorem). In addition,
 Regge theory, relates
different processes thanks to {\it factorization}, i.e.:
\begin{equation}
  \label{eq:factorization}
  \hbox{Im}\, F_{A+B\rightarrow A+B}(s,t)\simeq f_A(t) f_B(t) (s/\hat s )^{\alpha_R(t)},
\quad \hat s= (1\, \hbox{GeV})^2.
\end{equation}
Let us recall that total cross sections
are related to forward scattering amplitudes by:
$\sigma_{AB}=4\pi^ 2 \hbox{Im}\, F_{A+B\rightarrow A+B}(s,0)
/\lambda^{1/2}(s,m_A^2,m_B^2)$, with $\lambda(a,b,c)=a^2+b^2+c^2-2ab-2ac-2bc$.
The $ (s/\hat s )^{\alpha_R(t)}$ behavior depends on the Regge pole $R$,
whereas the $f_i(t)$ factors depend on the particles in the initial state.
It is then possible to obtain $\pi\pi$ Regge amplitudes 
from those of $\pi N$ and $NN$. The $(s,t)$
applicability range of the Regge formalism will be discussed below,
but QCD specifies that
$s>>\Lambda_{QCD}^2$ and $s>>|t|$. 

In the early seventies, when only phase shift analysis up to 2 GeV \cite{CERN-Munich}
were available for $\pi\pi$ scattering, and QCD was not fully developed, it was found that
certain crossing symmetry sum rules were not 
satisfied together with factorization \cite{Pennington:1974kp}, which, at that time,
implied $\sigma^\infty\simeq 15 \,$mb from $\pi N$ and $NN$ data. 
(In practice, $\infty$ means $\sim 20$ GeV, since above that energy
hadronic cross sections grow, as we will see below).
Thus,
it was suggested that factorization could be ``badly broken'', and that
$\sigma^\infty=6\pm5\,$mb.
Still, the same authors\cite{Martin:1977ff} remarked later that
$\sigma^\infty$ should be raised at least to 8.3 mb, and that
recent high energy measurements\cite{totaldata} of total $\pi\pi$ cross sections 
found $\sigma^{tot}_{\pi+\pi-}=13.5\pm2.5\,$mb at $32\,\hbox{GeV}^2$.
In the late seventies other authors\cite{Froggatt:hu} used the correct value in their analysis,
but when several other experiments in the late seventies confirmed this result, the interest
in Regge theory and pion scattering had already faded away.
 
In recent years, there is a renewed interest due to 
the implications for Chiral Perturbation Theory, through the precise
determination of the meson-meson scattering lengths, which requires
the use of dispersive integrals. Unaware of the last experiments\cite{totaldata},
all these studies \cite{Ananthanarayan:2000ht,Colangelo:2001df,disp}
have used a reanalysis\cite{Ananthanarayan:2000ht}
of the $\pi\pi$ Regge description with $\sigma^\infty=5\pm3\,$mb.

\section{Regge description of hadronic total cross sections including $\pi\pi$
scattering}

For the previous reasons we have recently presented \cite{Pelaez:2003ky} a Regge fit
able to describe 
$NN$,  $\pi^\pm N$, $K^\pm N$ and $\pi\pi$ from 
 $E_{kin}\simeq1 \hbox{GeV}$
up to about $\simeq16 \hbox{GeV}$.
We refer the readers to \cite{Pelaez:2003ky} for details.
Note that this new fit using $\pi\pi$ data  is compatible, but more precise
than a similar parametrization\cite{Pelaez:2003eh}, obtained from factorization
before we rediscovered the existing high energy data.
In Fig.1a, we show our fit to $\pi^\pm N$ and $(pp+p\bar p)/2$ 
and $(K^+ p+K^- p)/2$ cross sections, (these
combinations basically only depend on
the Pomeron and the $\rho$).
Here we use a very simple parametrization of the Pomeron with $\alpha_P(0)=1$,
although it is well known that hadronic cross sections grow at large $s$.
Nevertheless, as shown in Fig.1b,
with a slight modification of the Pomeron channel, to include a soft logarithm, 
the hadronic forward 
cross sections are described up to the multi-TeV region. We predict, for instance,
the total $pp$ cross section at LHC to be 
$\sigma^{tot}_{pp}(14)\hbox{TeV}=108\pm4\pm4\,$mb, in remarkable
agreement with the most recent Regge analysis \cite{Cudell:2002xe}: $111.5\pm1.2^{+4.1}_{-2.1}$.
Note also that, at low energies, Fig.1b overlaps with Fig.1a. 
\begin{figure}[h]
  \begin{center}
\includegraphics[scale=.75]{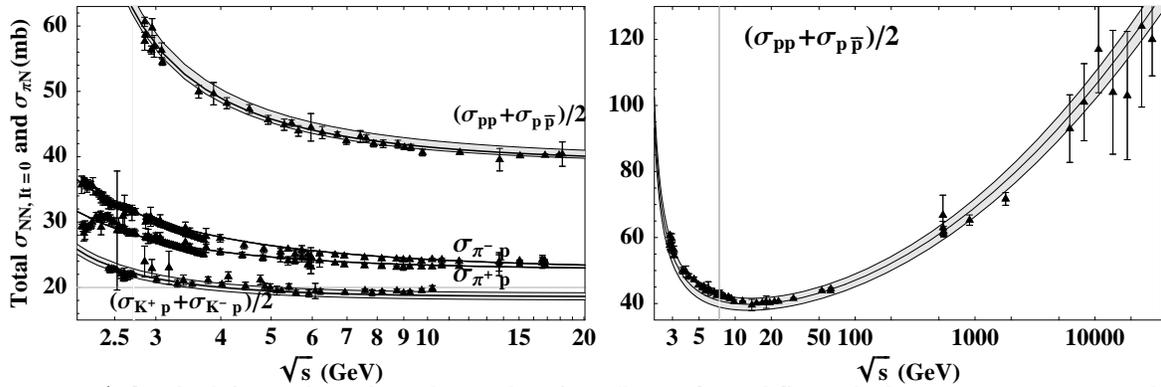}
\vspace{-.5cm}
\caption{\rm a) On the left panel we show the results of our
Regge fit to different hadronic cross sections, with a simple Pomeron
parametrization $\alpha_P(0)=1$, valid up to $E_{kin}\simeq16\,$GeV,
 and assuming factorization.
b) Adding a logarithmic term to the Pomeron the description can be easily extended
up to the multi TeV range.}
\end{center}
\end{figure}

However, the raise in hadronic forward cross sections 
is negligible up to 20 GeV, and is irrelevant for 
the integrals used in meson-meson
dispersive approaches. Hence, it is sufficient
to use the simplest Pomeron parametrization of Fig 1a.
assuming that factorization is also valid for $\pi\pi$ scattering
to obtain the results in Figs.2,
represented as solid lines. The thin gray bands around them
correspond to our error bars. We find a remarkable
agreement with the measured $\pi\pi$ cross sections above 2 GeV.
Between 1.4 and 2 GeV different sets of data are in conflict, note however
that our result falls between the different sets and that 
it matches with the points at 1.42 GeV from phase shift
analysis (the stars or the dotted lines). 
Finally, we find that all $\pi\pi$ amplitudes flatten around 20 GeV
and fall within 0.5 mb of an average value 
$\sigma^{tot}_{\pi\pi}(20 \mbox{GeV})=13.4\pm0.6\,$mb.
This value is totally dominated by the Pomeron and, assuming factorization,
is in very good agreement with the Regge parameterization \cite{Cudell:2001pn}
that will appear in the Review of Particle Properties 2004 \cite{PDG}, 
$\simeq12.3\pm0.3\,$mb.
For comparison
we show, as dashed lines, the results using the parameterizations in 
\cite{Ananthanarayan:2000ht} with their uncertainties (light gray bands). 
\begin{figure}[h]
  \begin{center}
\includegraphics[scale=.75]{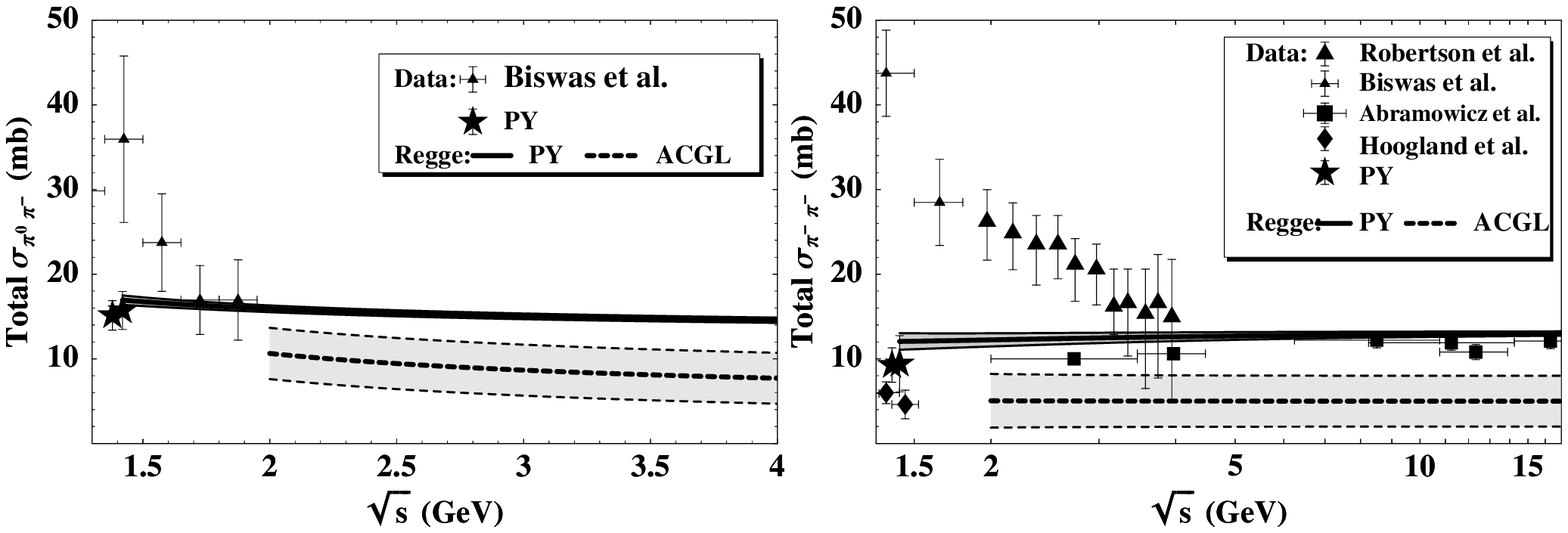}

\includegraphics[scale=.81]{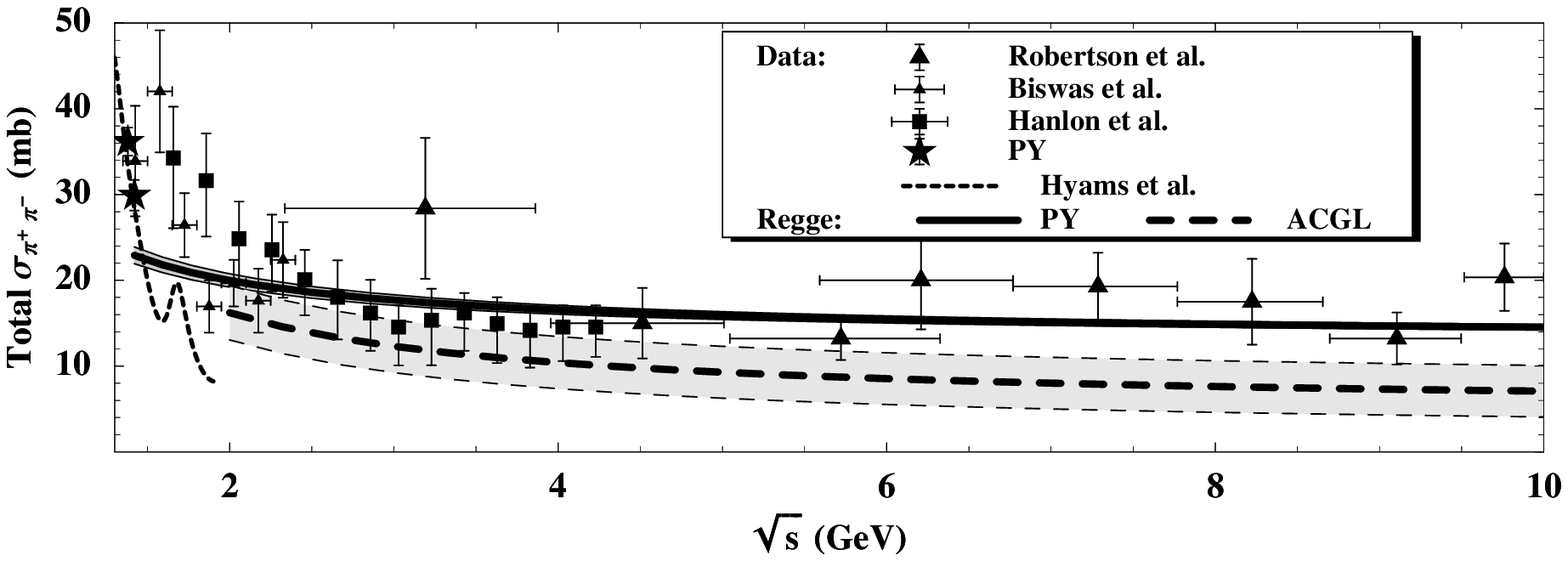}
\vspace{-.5cm}
\caption{\rm $\pi^0\pi^-$, $\pi^-\pi^-$
and $\pi^ +\pi^-$ total cross sections and data $^4$.The solid lines correspond to our
Regge fit assuming factorization. The dashed correspond lines to the parametrization $^6$ 
most frequently used in recent dispersive studies of $\pi\pi$ scattering.
The gray bands cover the uncertainties of each parametrization. The 
PY points have been obtained $^{10}$ from
parameterizations of phase shift analysis.}
\end{center}
\end{figure}

\section{$\pi\pi$ scattering and crossing sum rules}

We have also checked that our Regge description of $\pi\pi$
is consistent with crossing sum rules.
In particular, our 
parametrization without fitting \cite{Pelaez:2003ky}, 
satisfies the following two crossing sum rules:
\begin{eqnarray*}
\hspace*{-.8cm}&&\int_{4M^2_\pi}^\infty d\, s\,\Bigg\{
\frac{4\hbox{Im} F'^{I_s=0}(s,0)-10\hbox{Im} F'^{I_s=2}(s,0)}{s^2(s-4M^2_\pi)^2}
-6(3s-4m^2_\pi)\,
\frac{\hbox{Im} F'^{I_s=1}(s,0)-\hbox{Im} F^{I_s=)}(s,0)}{s^2(s-4M^2_\pi)^3}
\Bigg\}=0,\\
\hspace*{-.8cm}&&\int_{M^2_\pi}^\infty d\, s\,
\frac{\hbox{Im} F^{(I_t=1)}(s,4 M_\pi^2)-\hbox{Im} F^{(I_t=1)}(s,0)}{s^2}-
 \int_{M^2_\pi}^\infty d\, s\,\frac{8M^2_\pi[s-2M^2_\pi]}{s^2(s-4M^2_\pi)^2}
\hbox{Im} F^{(I_s=1)}(s,0)=0.
\end{eqnarray*}
where $F'$ means $\partial F/\partial cos \theta$, $\theta$ being the 
scattering angle. 

For these calculations we need the $t$ dependence 
of the Regge expressions, obtained from fits to the slopes of
differential cross sections (see \cite{Pelaez:2003ky} for detailed expressions). 
These fits are not unique in the literature, but
fortunately they agree numerically for small values of $t$. 
For the low $t$ we are interested in, $\sqrt{|t|}=0.28\,$GeV,
our $t$ behavior lies on the safe side \cite{Yndurain:2004aa}.

Both sum rules are well satisfied \cite{Pelaez:2003ky}
with the recent and most precise Regge parameters obtained from the fits in the
previous section. The first one is dominated by the Pomeron, and
 can be used to constrain the poorly known $I=2$ Regge exchange.
The second one is dominated at high energy by the $\rho$ and at low energy by the
$P$ wave, the best known. In this way one can fix with even greater precision
the Regge $\rho$ parameters, used to obtain Fig.2.

The old mismatches in these sum rules were due to the use of 
the CERN-Munich phase shift analysis \cite{CERN-Munich}
from $\sqrt{s}=1.42\,$GeV up to 2 GeV. They have bee represented by a dotted line
in Fig.2, where we can see that they are incompatible with measurements
of total cross sections. These data are also in conflict with many other considerations 
\cite{Pelaez:2003ky,Yndurain:2003jd}.

\section{Relevance for $\pi\pi$ scattering at low energies}

In a recent paper, we pointed out that the
$\pi\pi$ parameterizations given in \cite{Colangelo:2001df},
do not satisfy the Olsson sum rule by 2.5 standard deviations
and the $a_{00}$ and $a_{0+}$ Froissart Gribov dispersive representations
by about 4 to 5 standard deviations. 
Other dispersive analysis \cite{disp} do not have these conflicts because
their error bars are about a factor of three larger than those in  \cite{Colangelo:2001df}.
In \cite{Caprini:2003ta} it was argued
that this was due to a faulty Regge representation that violated crossing.
As shown here, our recent results \cite{Pelaez:2003ky} show that  this is not
the case, since our Regge formalism describes data and satisfies crossing sum rules.
The discrepancy remains.  

\section*{Acknowledgments}

This review is based on a set of works\cite{Pelaez:2003ky,Pelaez:2003eh} 
done in collaboration with prof. F.J.Yndurain to whom I thank
for his useful comments and suggestions.  Funds provided by the Spanish DGCICYT 
grants BFM2002-03218, BFM2000-1326 and BFM2002-01003
and EURIDICE contract HPRN-CT-2002-00311.

\end{document}